\nonstopmode

\newcommand{\be}{\begin{equation}}
\newcommand{\ee}{\end{equation}}
\newcommand{\bea}{\begin{eqnarray}}
\newcommand{\eea}{\end{eqnarray}}

\documentclass[twocolumn,superscriptaddress,groupedaddress,showpacs]{revtex4}
\usepackage{graphicx}
\usepackage{amsmath}
\usepackage{color}

\begin{document}
\title{Large Fluctuations and Singular Behavior of Nonequilibrium Systems}
\author{D.~Pinna}
\email{daniele.pinna@nyu.edu}
\affiliation{Department of Physics, New York University, New York, NY 10003, USA}
\author{A.D.~Kent}
\email{andy.kent@nyu.edu}
\affiliation{Department of Physics, New York University, New York, NY 10003, USA}
\author{D.L.~Stein}
\email{daniel.stein@nyu.edu}
\affiliation{Department of Physics, New York University, New York, NY 10003, USA}
\affiliation{Courant Institute of Mathematical Sciences, New York University, New
             York, NY 10012, USA} 
\affiliation{NYU-ECNU Institute of Physics at NYU Shanghai, 3663 Zhongshan Road North, Shanghai, 200062, China}
\date{\today}

\begin{abstract}
We present a general geometrical approach to the problem of escape from a metastable state in the presence of noise. The accompanying analysis leads to a simple condition, based on the norm of the drift field, for determining whether caustic singularities alter the escape trajectories when detailed balance is absent. We apply our methods to systems lacking detailed balance, including a nanomagnet with a biaxial magnetic  anisotropy and subject to a spin transfer torque. The approach described within allows determination of the regions of experimental parameter space that admit caustics.
\end{abstract}

\maketitle

{\it Introduction.\/} A noisy dynamical system 
will occasionally experience large fluctuations that can dramatically
alter its state. These fluctuations are responsible for a wide variety of interesting behaviors, including stochastic
resonance~\cite{stochres,Loffstedt}, transport via Brownian
ratchets~\cite{ratchets,Millonas}, logarithmic susceptibility in driven non-adiabatic systems~\cite{SmelyanskiyPRL}, and Brownian vortices~\cite{Bo1,Bo2}. In the limit of weak noise, the system's dynamical response is determined by its optimal paths, i.e., the paths along which it moves with maximum probability~\cite{FW,DykRev}.

When the system's stationary probability distribution lacks detailed
balance, the optimal paths can exhibit unusual behavior. In particular, singularities known as {\it caustics\/} can develop in the
action of the quasistationary density~\cite{Millonas,Dyk3,Maier1992,Maier1993b,Kogan2011} resulting from folds and cusps in the projection of the Lagrangian manifold of escape trajectories onto the original space of the dynamical
variables~\cite{Dyk3}. Such singularities cannot occur in the presence of
detailed balance, but in its absence their presence can significantly alter the behavior
of noise-induced escape from a static metastable state~\cite{Dyk3,Maier1993b} or a limit cycle~\cite{SmelyanskiyPRE, Maier1996}. 
Optimal trajectories avoid caustics~\cite{Dyk3}, so appearance of
a caustic in the vicinity of a stable or saddle point can dramatically alter the escape behavior~\cite{Maier1993b}. 

In this Letter, we reformulate the escape problem and in so doing determine 
conditions under which singularities in the optimal escape
trajectories can appear. This leads to a new
approach, based on a simple feature of the deterministic (i.e., zero-noise) dynamics, toward determining the presence and behavior of caustics. We apply
our approach, first to a previously studied system~\cite{Maier1993b} in which
caustics are known to dramatically affect the escape dynamics, and then to
a system not studied from this perspective, namely magnetic~reversal in a biaxial~nanomagnet subject to both thermal noise and spin transfer torque. For the latter system we determine the experimental parameter ranges for which caustics are likely
(and unlikely) to occur.

{\it Escape in a noisy dynamical system.\/} Consider an overdamped particle with position vector~${\mathbf{x}}(t)$ in an
$n$-dimensional space.  If the particle is subject to both deterministic and random forces, its time
evolution in the general case is described by the Langevin equation 
\be
\label{eq:Langevin}
\dot{\mathbf{x}}=\mathbf{F}(\mathbf{x})+\sqrt{2\epsilon}\mathbf{\hat{H}}\cdot\dot{\mathbf{W}}\, ,
\ee 
where $\mathbf{F}(\mathbf{x})$ denotes the drift field,
$\dot{\mathbf{W}}$ represents a white noise process, and the tensor
$\mathbf{\hat{H}}(\mathbf{x})$ and scalar $\epsilon$ characterize the
noise anisotropy and strength, respectively. We take $\epsilon$ to be
sufficiently small so the timescale of a successful escape is much
longer than that of a single excursion from the stable point.

Diagonalizing~(\ref{eq:Langevin}), we let $\mathbf{\hat{G}}$ be the
diagonal matrix with $\hat{G}_{ii}=g_i^2$ corresponding to $\mathbf{\hat{H}}$; the associated drift field vector then
has components~$f_i=F_i/g_i^2$. The quasi-stationary distribution~$\rho(\mathbf{x})\propto\exp[-S(\mathbf{x})/2\epsilon]$ is given to
leading order in the zero-noise limit~\cite{FW} by the solution of the
variational problem $S=\min\int_{-\infty}^\infty
L_{FW}\;\mathrm{d}t$, with the {\it Friedlin-Wentzell}~(FW) Lagrangian
and associated Hamiltonian given by 
\bea
\label{eq:Lag}
L_{FW}&=&\frac{1}{2}(\dot{\mathbf{x}}-\mathbf{F})\cdot\hat{G}^{-1}\cdot(\dot{\mathbf{x}}-\mathbf{F})\\
\label{eq:Ham}
H_{FW}&=&\frac{1}{2}\left[(\mathbf{p}+\mathbf{f})\cdot\hat{G}\cdot(\mathbf{p}+\mathbf{f})-\mathbf{f}\cdot\hat{G}\cdot\mathbf{f}\right]\, .
\eea 
Where $\mathbf{p}$ is the FW canonical momentum with elements~$p_i=\partial_{\dot{x}_i}L_{FW}$. 

Suppose now that the drift field contains one or more stable fixed points, and that the
system's initial state lies within the basin of attraction of one of these. It can be seen
by inspection of~(\ref{eq:Ham}) that the fluctuational dynamics can
be mapped onto the problem of a particle with unit electrical charge
moving on a Riemannian manifold with (positive definite) metric tensor
$\mathbf{\hat{G}}$, under the combined influence of a magnetic vector
potential $\mathbf{A}=-\mathbf{f}$~\cite{Dykman1983} and electric scalar potential
$\phi=-\mathbf{f}\cdot\hat{G}\cdot\mathbf{f}=-|\mathbf{f}|^2_{\mathbf{\hat{G}}}$~\cite{note}.
Hamilton's equations of motion are then: 
\bea
\label{eq:FullMotiona}
\dot{\mathbf{x}}&=&\hat{G}\cdot\mathbf{p}+\mathbf{F}=\hat{G}\cdot(\mathbf{p}+\mathbf{f})\\
\label{eq:FullMotionb}
\dot{\mathbf{p}}&=&-\nabla_{\mathbf{x}}H_{FW}\, .  
\eea 
The optimal escape paths are zero energy trajectories, that is, those
along which the Hamiltonian~(\ref{eq:Ham}) vanishes~(see, for example,~\cite{Maier2}). For
any state vector $\mathbf{x}$, the set of momenta satisfying the
zero-energy condition $H_{FW}=0$ can be seen to describe an
$n$-dimensional ellipse in momentum space centered at
$\mathbf{p}=-\mathbf{f}$ and with axes
$a_i=\sqrt{\mathbf{f}\cdot\hat{G}\cdot\mathbf{f}/g_i^2}$.

{\it The momentum ellipse.\/} For simplicity we now confine our
considerations to two dimensions; the procedure
presented below is straightforwardly generalized to higher
dimensions by parametrizing an~$n$-dimensional
ellipse. The $2D$~momentum ellipse can be parametrized as:
\bea
\label{eq:P_parama}
p_x(\gamma)&=&\frac{|\mathbf{f}|_{\hat{G}}}{g_x}\cos\gamma-f_x\\
\label{eq:P_paramb}
p_y(\gamma)&=&\frac{|\mathbf{f}|_{\hat{G}}}{g_y}\sin\gamma-f_y\, .
\eea

\begin{figure}
	\begin{center}
	\centerline{\includegraphics [clip, width=70mm, angle=0]{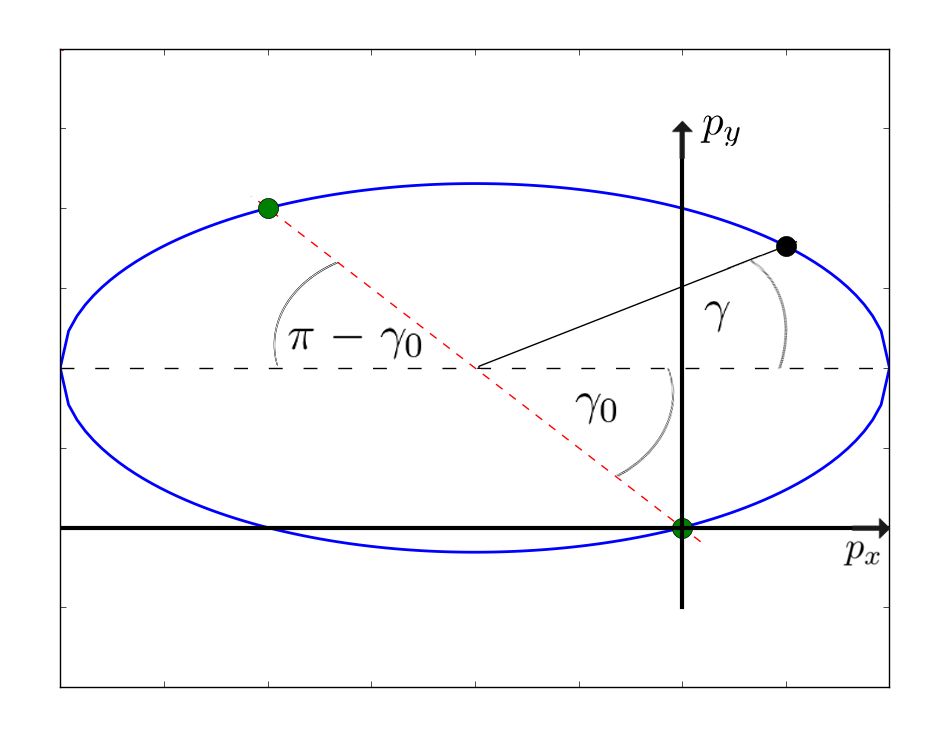}}
	 \vskip -4mm
	\caption{Diagram of momentum ellipse parametrized by $\gamma$. $\gamma_0$ and $\pi+\gamma_0$ correspond to instanton and anti-instanton solutions respectively.}
	\label{fig:momellipse}
	\vskip -2mm
	\end{center}
\end{figure}

This momentum ellipse defines all possible least-action motions
accessible to the particle at  $\mathbf{x}$ traveling along an escape trajectory. The usual anti-instanton trajectories ($\bar{p}=0$)
characterizing zero-noise drift correspond to
$\gamma=\gamma_0\equiv-\arctan(\frac{f_y}{f_x}\frac{g_y}{g_x})$;
instanton trajectories for systems satisfying detailed balance
correspond to motion antiparallel to the drift fields
($\bar{p}=-2\bar{f}$)~\cite{FW,Maier2,Marder}, which here corresponds to
$\gamma=\pi+\gamma_0$. Fig.~\ref{fig:momellipse} shows a typical
diagram of such a momentum ellipse.

Because the angle $\gamma$ parametrizes momentum at each point in
configuration space, it is convenient to express the equations of motion of 
the particle trajectories solely as a function
of $\mathbf{x}$ and $\gamma$. Substituting~(\ref{eq:P_parama}) and~(\ref{eq:P_paramb}) into the
Hamiltonian equations of motion~(\ref{eq:FullMotiona}) and~(\ref{eq:FullMotionb}) , we find

\bea
\label{eq:gammaxdot}
\dot{x}&=&g_x|\mathbf{f}|_{\mathbf{\hat{G}}}\cos\gamma\nonumber\\
\dot{y}&=&g_y|\mathbf{f}|_{\mathbf{\hat{G}}}\sin\gamma.
\eea
The role of $\gamma$ in characterizing the direction of escape is
apparent. The slope of the escape trajectory is found by dividing the second of
Eqs.~(\ref{eq:gammaxdot}) by the first to yield
\be
\label{eq:slope}
\partial y/\partial x=(g_y/g_x)\tan\gamma\, .  
\ee 

{\it The nature of fluctuational trajectories}. Eqs.~(\ref{eq:gammaxdot}) show that thermally driven dynamics evolve at the rate of the deterministic dynamics rescaled by the noise-induced metric, i.e.~$|\dot{x}|_{\mathbf{\hat{G}}^{-1}}^2=|\mathbf{f}|_{\mathbf{\hat{G}}}^2=|\mathbf{F}|_{\mathbf{\hat{G}}^{-1}}^2$. The two rates become trivially identical for isotropic noise ($\mathbf{\hat{G}}\equiv\mathbf{\hat{1}}$) where detailed balance is satisfied, because the instanton $\dot{\mathbf{x}}=-\mathbf{f}$ and anti-instanton
$\dot{\mathbf{x}}=\mathbf{f}$ trajectories are simply sign-reversed.  It
is somewhat surprising, however, to see that it holds more generally. This result can alternatively be derived by writing down the effective Lorentz dynamics for a charged particle traveling in both electric and
magnetic fields 
\be
\label{eq:Lorentz}
\ddot{\mathbf{x}}=\nabla|\mathbf{f}|^2-\dot{\mathbf{x}}\times(\nabla\times\mathbf{f})
\ee
and noting that upon multiplying by $\dot{\mathbf{x}}$ one obtains
$\partial_t|\dot{\mathbf{x}}|^2=\partial_t|\mathbf{f}|^2$, again
implying that the dynamical speed of a particle moving under the
influence of noise is equal to the norm of the zero-noise drift
field. This notion has been employed in the
literature~\cite{Heymann2008} to construct an efficient
numerical scheme (the {\it gMAM} method) capable of computing
transition pathways via geometric minimization of the FW action, improving on the older 
String~method~\cite{Weinan2002,Ren2007}.  

Using these results, the Friedlin-Wentzel~Lagrangian can be rewritten as:
\be
\label{eq:lag}
\begin{split}
L_{FW}=\frac{1}{2}|\dot{\mathbf{x}}-\mathbf{F}|_{\mathbf{\hat{G}}^{-1}}^2&=\frac{1}{2}\left[|\dot{\mathbf{x}}|_{\mathbf{\hat{G}}^{-1}}^2+|\mathbf{F}|_{\mathbf{\hat{G}}^{-1}}^2-2\dot{\mathbf{x}}\cdot\mathbf{\hat{G}^{-1}}\cdot\mathbf{F}\right]\\
&=|\mathbf{f}|_{\mathbf{\hat{G}}}^2-\dot{\mathbf{x}}\cdot\mathbf{f}\\
&=|\mathbf{f}|_{\mathbf{\hat{G}}}^2\left(1-\cos\Psi\right)
\end{split}
\ee
where we have employed the identities
$\mathbf{F}=\mathbf{\hat{G}}\cdot\mathbf{f}$ and
$\dot{\mathbf{x}}\cdot\mathbf{\hat{G}}^{-1}\cdot\dot{\mathbf{x}}=|\mathbf{f}|_{\mathbf{\hat{G}}}^2$,
and defined
\be
\label{eq:psi}
\Psi\equiv\mathrm{arccos}\left(\frac{\dot{\mathbf{x}}\cdot\mathbf{f}}{|\mathbf{f}|_{\mathbf{\hat{G}}}^2}\right)
\ee as the angle between the instantaneous escape velocity
$\dot{\mathbf{x}}$ and the deterministic drift field~$\mathbf{f}$ at
$\mathbf{x}$.  Eq.~(\ref{eq:psi}) shows that, as long as $\Psi$ does
not vary too much over the course of the escape trajectory, the
effective action $S(\mathbf{x})=\int_{-\infty}^{\infty}L_{FW}dt$ will
be dominated by the behavior of $|\mathbf{f}|_{\mathbf{\hat{G}}}$,
implying that the norm of the drift field alone captures much of the
structure of the system's action. This is the central result of the paper.

To characterize better the relative importance of $|\mathbf{f}|$ vs. $\Psi$ on the escape dynamics, consider a closed fluctuational path ${\cal C}$ that moves from a stable fixed point $\mathbf{x_S}$ to some other point $\mathbf{x_0}$ within its attractive basin, subsequently returning (along the anti-instanton trajectory $\dot{\mathbf{x}}=\mathbf{f}$) to the fixed point. Using $\mathbf{\Omega_{\cal C}}$ to denote the surface enclosed by~$\mathbf{\cal C}$, we find 
\begin{equation}
\label{eq:LagPhase}
S_{\cal C}=\oint_{\cal C} \left(|\mathbf{f}|_{\mathbf{\hat{G}}}^2-\dot{\mathbf{x}}\cdot\mathbf{f}\right)\mathrm{d}t=\oint_{\cal C}\mathrm{d}s|\mathbf{f}|_{\mathbf{\hat{G}}}-\int\mathrm{d}\mathbf{\Omega_{\cal C}}\cdot\left(\nabla\times\mathbf{f}\right)\, .
\end{equation}
In the second equality we changed integration variables from $t$ to $s$, where $s$ denotes distance traveled along the trajectory, and used the earlier result that $|\dot{\mathbf{x}}|=|\mathbf{f}|$ for all systems. We also used Stokes'~theorem to rewrite the effect of the geometrical phase due to $\Psi$ in terms of its equivalent surface integral. This equation shows how the presence of a nongradient field (which possesses nonzero curl) contributes to the action, and therefore helps determine the trajectory, of an optimal path. While enclosing a larger `magnetic flux'~$\int\mathrm{d}\mathbf{\Omega_{\cal C}}\cdot(\nabla\times\mathbf{f})$ acts to reduce the path's action (where the flux term is positive), doing so can also move the particle into regions of configuration space with larger~$|\mathbf{f}|$, which acts to increase the action. The least action trajectory  therefore optimizes the relative balance between these two terms. 

There are two immediate consequences of~(\ref{eq:LagPhase}). The first is that in gradient systems, where $\mathbf{f}=-\mathbf{\nabla}_{\mathbf{x}}{U}$ for some smooth potential function~$U(\mathbf{x})$ (so $\nabla\times\mathbf{f}=0$), and where the instanton trajectory satisfies $\dot{\mathbf{x}}=-\mathbf{f}$, we recover the well-known~action~(see, for example,~\cite{Maier2}) for the entire trajectory:~$S_{\cal C}=2\Big[U(\mathbf{x_0})-U(\mathbf{x_S})\Big]$. The second, more general consequence is that any sudden change in $|\mathbf{f}|$, as system parameters are varied, can dramatically alter the balance between the drift norm and the magnetic flux, and hence the structure of the optimal paths.

{\it Applications.\/} It has been
observed~\cite{Maier1993b,Dyk3} that noisy systems with nongradient
deterministic dynamics containing tunable parameters can dramatically change
their escape behaviors, in a way reminiscent of a broken symmetry transition, when certain critical parameter
thresholds are crossed.  Consider, for example, the following
well-studied system~\cite{Maier1993b} with a tunable parameter $\alpha$:
\bea
\dot{x}&=&x(1-x^2-\alpha y^2)+\sqrt{\epsilon}\dot W_x\\
\dot{y}&=&-y(1+x^2)+\sqrt{\epsilon}\dot W_y,
\eea
where the additive white noise is isotropic and the drift field is
nongradient for all $\alpha\ne 1$. For any $\alpha$, the system has two
stable fixed points at $(\pm 1,0)$ and one hyperbolic (saddle) fixed
point at $(0,0)$. We consider the optimal escape trajectory from
$(1,0)$ to $(0,0)$, which flows along the $x$-axis for $\alpha<4$. Above $\alpha=4$, however, 
caustics appear in the basin of attraction
of the stable point~\cite{Maier1993b}, focusing to a point on the $x$-axis between 0 and
1, as shown in Fig.~2. As a consequence, the optimal escape trajectory bifurcates into two off-axis trajectories. 
\begin{figure}
	\begin{center}
	\centerline{\includegraphics [clip,width=70mm, angle=0]{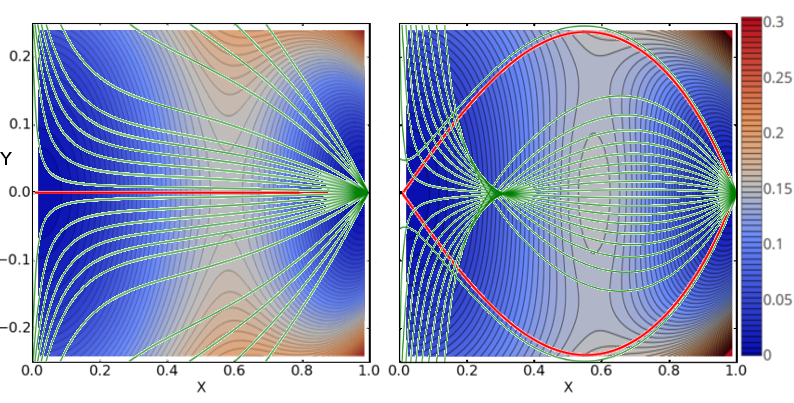}}
	\vskip -4mm
	\caption{Contour plot of the norm of the drift field from~\cite{Maier1993b}, for $\alpha=3$~(left) and $\alpha=5$~(right). Optimal escape trajectories are shown in red, and other instanton trajectories in green. For
          $\alpha=3$, instanton trajectories do not cross, and the escape trajectory lies along the $x$-axis. Correspondingly, two global minima in the norm of the drift field are present at the unstable
          $(0,0)$ and stable $(1,0)$ equilibria, respectively, along
          with a saddle close to the midpoint of the x-axis. For
          $\alpha=5$, instanton trajectories cross, indicating the presence of a caustic (not shown). Correspondingly, the saddle in the norm of the drift field for $\alpha=3$ is now a local maximum, with two new local minima appearing off the
          $x$-axis. Consequently, there are now two symmetrical off-axis optimal escape trajectories
          that follow the new off-axis minima in the
          norm of the drift field (and in so doing avoid the caustic).}
\label{fig:Stein}
    \vskip -2mm
	\end{center}
\end{figure}
Analysis of the norm of the drift field reveals that in transitioning
across the critical threshold $\alpha_C=4$, the structure of the
extrema of $\mathbf{f}$ changes abruptly. For~$\alpha<4$, the norm
exhibits two global minima at $(1,0)$ and $(0,0)$, along with a saddle
near the midpoint of the $x$-axis. For $\alpha>4$, however, two local minima appear 
symmetrically displaced off
the $x$-axis, with the previous on-axis saddle now a local maximum~(Fig.~\ref{fig:Stein}).  
According to~(\ref{eq:lag}), these new minima lower the effective action of off-axis escape trajectories, in accordance with
observations. Deviation of the escape trajectories from the exact minima are due to the `magnetic flux' term in~(\ref{eq:LagPhase}).

A more interesting application is to a physically relevant model
lacking detailed balance: the {stochastic Landau-Lifshitz-Gilbert-Slonczewski}~(sLLGS) 
equation governing the evolution of a unit magnetization vector subject to a combination of
both gradient and non-gradient torques. This equation reads:
\begin{equation}
\label{eq:Langevindynamics}
\dot{m}_i=A_i(\mathbf{m})+B_{ik}(\mathbf{m})\circ H_{th,k},
\end{equation}
where the drift vector $\mathbf{A}(\mathbf{m})$ and diffusion matrix $\hat{\mathbf{B}}(\mathbf{m})$ are given by
\begin{eqnarray}
\label{eq:sLLGS}
\mathbf{A}(\mathbf{m})&=&\mathbf{m}\times\mathbf{h}_{\mathrm{eff}}-\alpha\mathbf{m}\times\left(\mathbf{m}\times\mathbf{h}_{\mathrm{eff}}\right)\nonumber\\
&-&\alpha I\mathbf{m}\times\left(\mathbf{m}\times\mathbf{\hat{n}}_p\right),\\
B_{ik}(\mathbf{m})&=&\sqrt{C}[-\epsilon_{ijk}m_j-\alpha(m_i m_k - \delta_{ik})]
\end{eqnarray}
and the equation is interpreted in the Stratonovich sense. The first term in~$\mathbf{A}(\mathbf{m})$ corresponds to
magnetization precession about a local magnetic field, with
$\mathbf{h}_{\mathrm{eff}}=-\nabla_\mathbf{m} \epsilon(\mathbf{m})$ a conservative
vector field. Here  $\epsilon(\mathbf{m})$ is the energy landscape
of the magnetic system under study.  The second term in~$\mathbf{A}(\mathbf{m})$ is a phenomenological damping term, with
the damping constant~$\alpha$ typically~$\sim O(10^{-2})$. The third is a nongradient term
corresponding to spin-angular momentum per unit time injected via a current $I$ into
the macrospin along an arbitrary polarization direction
$\mathbf{\hat{n}}_p$~\cite{notealpha}. Although the diffusion
matrix~$\hat{\mathbf{B}}(\mathbf{m})$ (with $C$ the diffusion
constant) appears state-dependent, it can be shown (e.g., by rewriting the dynamics in
spherical coordinates) to correspond to isotropic, state-independent noise.

In the absence of applied currents ($I=0$), the fluctuational trajectories are determined by the energy landscape~$\epsilon(\mathbf{m})$
and do not cross.   In the presence of a nonvanishing current $I$, however, detailed balance is absent, and it therefore becomes important to determine how this new feature may --- or may not --- alter the escape dynamics. The 
methods developed above allow us to analyze this problem by
examining the norm of the total drift field governing the macrospin
dynamics.  To lowest order in~$\alpha$, it is
\begin{equation}
\label{eq:magnorm}
|\mathbf{A}(\mathbf{m})|^2=|\mathbf{m}\times\mathbf{h}_{\mathrm{eff}}|^2+2\alpha I\,\mathbf{\hat{n}}_p\cdot(\mathbf{m}\times\mathbf{h}_{\mathrm{eff}})+O(\alpha^2)\, .
\end{equation}

We wish to determine under which conditions the conservative precessional contribution dominates the
nongradient contribution. When this occurs, the extrema of $|\mathbf{A}(\mathbf{m})|^2$ corresponding to $I=0$ do not
change significantly when $I>0$. By the approach developed above, this implies that the most probable escape
paths should not differ significantly from the reverse-drift instanton paths of the purely gradient case.

We consider for simplicity the case of a biaxial macrospin subject to the energy landscape 
\begin{equation}
\label{eq:biaxial}
\epsilon(\mathbf{m})=Dm_x^2-m_z^2\, ,
\end{equation}
defined on the surface of the unit sphere ($|\mathbf{m}|=1$). Here $D\ge 0$ is the ratio 
of hard axis to easy axis anisotropies~\cite{Pinna2013,PSK14}. 

The calculation analyzing the relative contributions of the gradient and nongradient terms in~(\ref{eq:magnorm})
is done in detail in the Supplementary Notes~\cite{supp}. We find
that the presence of caustics is determined by the fixed tilt angle~$\omega$ that~$\mathbf{\hat{n}}_p$ makes with the~$z$-axis.
For any $D$, the escape dynamics are unaffected by caustics for a wide energy range in the small tilt case ($\tan\omega\ll\alpha^{-1}$), 
but in principle the large tilt case ($\tan\omega\gg\alpha^{-1}$) can show substantially different behavior.

\begin{figure*}
\centering
\setlength\fboxrule{0.0pt}
\fbox{\includegraphics[width=5.5in]{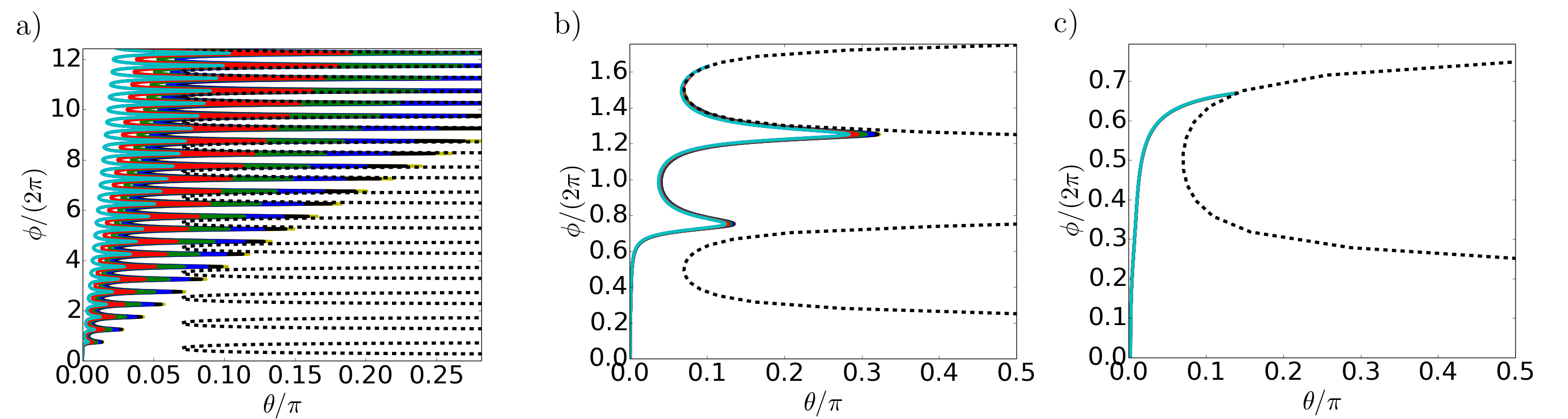}}
\caption{Escape trajectory for macrospin model with $\alpha=0.01$, $D=20$, $I=0.8\,I_C$ and varying tilts: (a) $\omega=0$, (b) $\omega=0.1\,\theta_C$, (c) $\omega=0.25\,\theta_C$. Different color trajectories correspond to different initial conditions of the FW~dynamics. Dashed black lines correspond to the $\epsilon=0$ separatrices. In all scenarios, escape trajectories never cross, indicating the absence of caustics.}
\label{fig:D20}
\end{figure*}

We tested these predictions by numerically integrating the FW~dynamics associated with the stochastic process~(\ref{eq:Langevindynamics}) for several different cases. Results for $D=0$ are shown in the Supplementary~Notes~\cite{supp}. The absence of caustics, for both zero and small tilt, is apparent. Here we present the more interesting case of nonvanishing $D$. Fig.~\ref{fig:D20} shows results for $D=20$, $I=0.8\,I_C$, and three tilts $\omega=0$, $0.1$ and $0.25\,\theta_C$, where $\theta_C=\arctan(1/\sqrt{D})$~\cite{supp} and $I_C$ is the critical applied current above which the entire $m_z>0$ region  becomes unstable, eliminating the bistability of the magnetic system. These tilt values (and even larger ones)  can be realized in current experiments on orthogonal spin-valve devices~\cite{Li}. The optimal trajectories change as tilt is increased, even though the exit point remains essentially unchanged. However, numerical results for larger tilts, up to the critical tilt, do not exhibit crossing of escape trajectories, suggesting that caustics have not formed near escape paths. The only major difference is the number of precessions the system undergoes before reaching the separatrix (i.e., the boundary of the domain of attraction of the stable fixed point, which here corresponds to $\epsilon=0$~\cite{supp}).

To summarize, an analysis of the norm of the drift field indicates that for most tilts, caustics do not appear within the escape region, and so --- perhaps surprisingly --- the loss of detailed balance does not qualitatively alter the escape dynamics.  However, on closer examination one finds that for any tilt, there remain regions --- where $m_z\approx 1$ or, separately, where $\epsilon\approx 0$ --- where this conclusion breaks down, because the precessional contribution to the drift norm vanishes at $m_z=1$ and at $\epsilon=0$. That is, two regions will always exist in the magnet's configuration space where the (nongradient) spin transfer torque term dominates the (gradient) precessional term. We compute the width of these regions in the Supplementary~Notes and show that they are sufficiently small that their effect on the escape dynamics is negligible.

In the case when the tilt is large ($\tan\omega\gg\alpha^{-1}$), however, the nongradient term dominates the gradient term in all of configuration space, and so caustics {\it may\/} appear. In order to determine whether they do requires a more lengthy analysis, in which the fixed point structure of the drift field norm must be analyzed to determine whether it changes. We defer such an analysis to future work.

\begin{acknowledgments}
This research was supported in part by U.S.~NSF Grant DMR-1309202. DLS thanks the John Simon Guggenheim Foundation for a fellowship that partially supported this research, and NYU~Paris and the Institut~Henri~Poincar\'e for their hospitality while part of this research was carried out. We thank Mark~Dykman and Guido D'Amico for many useful conversations and comments on the manuscript.
\end{acknowledgments}


\begin{thebibliography}{99}
\bibitem{note} Given a Riemannian metric~$\mathbf{\hat{Q}}$, we will denote the inner
  product of two vectors~$\mathbf{a}$ and~$\mathbf{b}$ as~$\langle\mathbf{a},\mathbf{b}\rangle_{\mathbf{\hat{Q}}}$. Analogously,
  the inner product on~$\mathbf{\hat{Q}}$ of a vector with itself will
  be written~$\langle\mathbf{a},\mathbf{a}\rangle_{\mathbf{\hat{Q}}}\equiv|\mathbf{a}|_{\mathbf{\hat{Q}}}^2$.
\bibitem{notealpha} The usual formulation of the spin-torque term does not include a factor of $\alpha$; it is not a dissipative term.
We include it here because it is convenient to define the current in terms of $\alpha$: $I\sim J/\alpha$, where $J$ is the
spin current pumped into the system per unit time. The reason for doing this is so the critical current $I_C$ for switching
(defined in the text two paragraphs below Eq.~(\ref{eq:biaxial})) is independent of $\alpha$: $I_C=J_C/\alpha=(2/\pi)\sqrt{D(D+1)}$.
\end{thebibliography}
\end{document}




\title{Supplemental Material for ``Large Fluctuations and Singular Behavior of Nonequilibrium Systems''}

\author{D.~Pinna}
\affiliation{Department of Physics, New York University, New York, NY 10003, USA}
\author{A.D.~Kent}
\affiliation{Department of Physics,,
New York University, New York, NY 10003}
\author{D.L.~Stein}
\affiliation{Department of Physics and Courant Institute of Mathematical Sciences,
New York University, New York, NY 10012}
\pacs{}
\maketitle

\section{S1: Relative contributions of gradient and nongradient terms to the drift field norm for a biaxial macrospin}

In this section we calculate the relative contributions of the gradient (precessional) and nongradient (spin torque) terms in
Eq.~(19) in the text. In the absence of applied currents and damping, the motion is purely
precessional. Focusing solely on this precessional motion, we can divide
the unit magnetization sphere into four disjoint dynamical regions, as
shown in~Fig.~1.
\begin{figure}[h]
	\begin{center}
	\centerline{\includegraphics [clip,width=2.5in, angle=0]{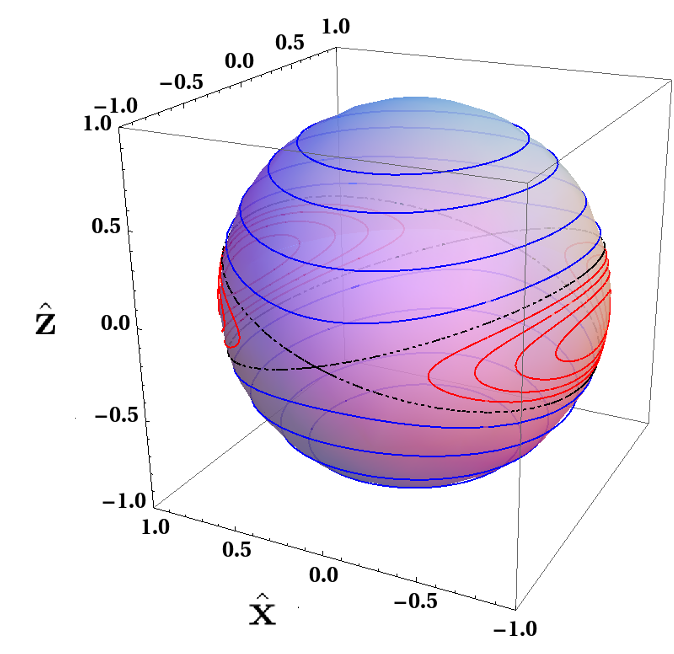}}
	\caption{Constant energy contours of the unit magnetization
          sphere. Red and blue contours correspond to $\epsilon<0$ and $\epsilon>0$ precessions respectively. Dashed black lines are the $\epsilon=0$ separatrices.}
	\label{fig:magsphere}
	\end{center}
\end{figure}
When magnetization energy is conserved, the surface of the sphere can be divided
into trajectories with $\epsilon>0$ and those with $\epsilon<0$. The
$\epsilon<0$ surface can be further subdivided into orbits with
$m_z>0$ and $m_z<0$, and the $\epsilon>0$ surface similarly subdivided
into orbits with $m_x>0$ and $m_x<0$. These regions are separated by a
common separatrix $m_x=\pm m_z/\sqrt{D}$, corresponding to a degenerate
$\epsilon=0$ dynamical state whose trajectory lies on a plane tilted at an angle~$\theta_C=\arctan(1/\sqrt{D})$ with respect to the $y$-$z$ plane.

However, the above description must be modified when damping and current are taken into account, since either will cause the magnetization energy~$\epsilon(\mathbf{m})$ to change over time.
Using~(16)-(18) from the text and the definition~$\mathbf{h}_{\mathrm{eff}}=-\nabla_\mathbf{m} \epsilon(\mathbf{m})$,
we find
\begin{equation}
\label{eq:edot}
\dot{\epsilon}=\nabla_{\mathbf{m}}\epsilon\cdot\dot{\mathbf{m}}=-\alpha\left[\mathbf{m}\times\left(\mathbf{h}_{\mathrm{eff}}+I\mathbf{\hat{n}}_p\right)\right]\cdot\left(\mathbf{m}\times\mathbf{h}_{\mathrm{eff}}\right)\, .
\end{equation}
Several qualitative conclusions can be drawn from~(\ref{eq:edot}). 
When $I=0$, dissipation relaxes the system to one of the two
negative energy basins in an effort to align the magnetization with
the $z$-axis. The addition of current (which we will conventionally
choose to be negative, $I<0$) leads to a progressive
destabilization of the $m_z>0$ negative energy basin in favor of that
at $m_z<0$. More generally, the presence of current alters the boundaries of the attractive stable
basins, even though a proper energy function
that includes the spin-torque terms cannot be written. However,
one can show that  there exists a critical value $D_0\approx 5.09$
that separates two distinctly different qualitative
regimes~\cite{Pinna2013}. In the $D<D_0$ case, the current
alters the $\epsilon=0$ separatrix boundary, while in the $D>D_0$
case, the $\epsilon=0$ boundary remains intact while the stable fixed
points are modified by the current's action. Furthermore, above a critical
current $I_C$ (to be discussed further below) the entire $m_z>0$ region 
becomes unstable, eliminating the bistability of the magnetic system. 

We want to determine under which conditions the nongradient current term can 
alter the fixed point structure of the norm of the drift field.  Because of the smallness of $\alpha$, energy changes slowly, so one can still gain useful information by
studying the norm of the drift along a constant negative energy trajectory. Along each energy level set $\epsilon$ one can compute the maxima and
minima of the two terms appearing on the RHS of~(19). This allows us to determine the range of
experimental parameters for which the precessional term dominates, and therefore when effects due to nongradient terms can safely be
ignored.  Using~(20) and the definition of $\mathbf{h}_{\mathrm{eff}}$, we find the minimum of the precessional term along a particular
constant energy contour to be:
\begin{equation}
\label{eq:min}
\mathrm{min}\{|\mathbf{m}\times\mathbf{h}_{\mathrm{eff}}|^2\}=4\,\mathrm{min}\{\epsilon(D-\epsilon)+(D+1)m_z^2\}=4|\epsilon|(1-|\epsilon|)\,, 
\end{equation}
where we used the fact that $\mathrm{min}\{m_z^2\}=-\epsilon$ along a constant negative energy contour. 

We next consider the maximum of the nongradient current term in~(19) and, for definiteness, 
we limit ourselves to cases where $\mathbf{\hat{n}}_p$ lies in the $x$-$z$ plane. Let~$\omega$ 
denote the fixed tilt angle that~$\mathbf{\hat{n}}_p$ makes with the~$z$-axis. Then for small~$\alpha$ 
the critical currents are~$I_C=(2/\pi)\sqrt{D(D+1)}/\cos\omega$  when $D>D_0$ and~$I_C=(D+2)/2\cos\omega$ when $D<D_0$~\cite{Pinna2013}. 
Using these results, we find
\begin{eqnarray}
\label{eq:max}
  \mathrm{max}\{2\alpha I\mathbf{\hat{n}}_p\cdot(\mathbf{m}\times\mathbf{h}_{\mathrm{eff}})\}&=&(4\alpha I_C\cos\omega)\mathrm{max}\{m_y\left(Dm_x+m_z\tan\omega\right)\}\nonumber\\
  &=&\begin{cases}
	4\alpha\sqrt{|\epsilon|(1-|\epsilon|)}\tan\omega, & \text{if $D=0$}.\\
	\frac{8\alpha}{\pi}\sqrt{D(D+1)(D+|\epsilon|)|\epsilon|}\,Q(\omega), & \text{if $D>D_0$}.\\
    	2\alpha(D+2)\sqrt{(D+|\epsilon|)|\epsilon|}\,Q(\omega), & \text{if $D<D_0$}.
  \end{cases}
\end{eqnarray}
where 
\begin{equation}
\label{eq:Q}
Q(\omega)=\mathrm{max_{m_z}}\left\{\sqrt{1-\frac{(D+1)m_z^2}{D+|\epsilon|}}\left[\sqrt{\frac{m_z^2}{|\epsilon|}-1}+\frac{\tan\omega}{\sqrt{D}}\frac{m_z}{\sqrt{|\epsilon|}}\right]\right\},
\end{equation}
and we set the current equal to $I_C$ in each case to obtain an upper bound. 

The above equations reveal a quantitative difference between the small tilt ($\tan\omega\ll \alpha^{-1}$) and large tilt ($\tan\omega\gg \alpha^{-1}$) cases.
Consider first the case~$D=0$.  By~(\ref{eq:min}) and~(\ref{eq:max}),
the precessional term dominates the nongradient term when
$\epsilon(1-\epsilon)\gg\alpha^2\tan^2\omega$.  For the small tilt case, as long as the magnetization energy is not too close to the separatrix ($\epsilon=0$) or
the $m_z=1$~pole ($|\epsilon|=1$), the precessional term dominates and no caustics will be present. More precisely, the precessional term dominates
within the energy range $\alpha^2\tan^2\omega<|\epsilon|<1-\alpha^2\tan^2\omega$. When the tilt is larger, on the other hand, the precessional term dominates
for most parameter values, and the presence of caustics {\it may\/} affect the reversal dynamics. Whether they do requires a detailed analysis of the particular case of interest. 

This conclusion remains qualitatively unchanged when $D>D_0$, with the main quantitative change being that the energy range in which caustics cannot appear 
in the small tilt case is now $D\alpha\tan\omega<|\epsilon|<1-D\alpha\tan\omega$. For $D<D_0$, we need to further subdivide into the $D>|\epsilon|$ and $D<|\epsilon|$ cases.
The latter smoothly goes to the $D=0$ case as $D$ decreases, while the former gives the same result as the $D>D_0$ case. Our overall conclusion is that for any $D$ the escape dynamics
are largely unaffected for a wide energy range in the small tilt case, but in principle the large tilt case can show substantially different behavior.

\bigskip
\bigskip

\section{S2: Analytical solution of the $\omega=0$ case}

In this section we analytically solve for the Hamiltonian dynamics arising from the FW~action for the case of $\omega=0$, with arbitrary applied current.
In this case, the dynamics
become effectively one-dimensional and the dynamical equations can be exactly solved.
The FW~Lagrangian for the thermally activated dynamics of the macrospin, expressed in spherical coordinates ($\theta$,$\phi$), is:

\begin{equation}
L_{FW}=\frac{1}{2}\left[(\dot{\theta}-f_{\theta})^2+\sin^2(\theta)(\dot{\phi}-f_{\phi})^2\right].
\end{equation}
Passing to the Hamiltonian formalism we have:
\begin{equation}
\label{sphEn}
H_{FW}=\frac{1}{2}\left[p_{\theta}^2+\frac{p_{\phi}^2}{\sin^2(\theta)}\right]+p_{\theta}f_{\theta}+p_{\phi}f_{\phi},
\end{equation}
where $p_{\theta}$ and $p_{\phi}$ are the conjugate momenta of $\theta$ and $\phi$.

Hamilton's equations are then  
\begin{eqnarray}
\dot{\theta}&=&p_{\theta}+f_{\theta}\\
\dot{\phi}&=&\frac{p_{\phi}}{\sin^2\theta}+f_{\phi}\\
\dot{p_{\theta}}&=&\frac{\cos\theta}{\sin^3\theta}p_{\phi}^2-p_{\theta}\partial_{\theta}f_{\theta}-p_{\phi}\partial_{\theta}f_{\phi}\\
\dot{p_{\phi}}&=&-p_{\theta}\partial_{\phi}f_{\theta}-p_{\phi}\partial_{\phi}f_{\phi},
\end{eqnarray}
where so far all results are completely general and extensible to models of any complexity (anisotropy ratio $D\neq0$, positive tilts between axes, etc.). For a macrospin with both~$D=0$ and~$\omega=0$, the drift vector field of the dynamics (expressed in terms of $\theta$,$\phi$) can be obtained by rewriting Eqs.~(16)-(18) from the text in spherical coordinates. One then obtains
\begin{eqnarray}
\label{eq:macrodrift}
f_{\theta}&=&-\alpha(I+\cos\theta)\sin\theta\\
f_{\phi}&=&-\cos\theta.
\end{eqnarray}
Because~$\phi$ is cyclic, $p_{\phi}=c$ with $c$ constant. For a zero-energy trajectory of~(\ref{sphEn}), $p_{\theta}$ is given by:
\begin{equation}
\label{eq:ptheta}
p_{\theta}=-f_{\theta}\left[1\pm\sqrt{1-c\left(2\frac{f_{\phi}}{f_{\theta}}+\frac{\alpha c}{\xi f_{\theta}\sin^2\theta}\right)}\right].
\end{equation}
This solution is physically valid as long as the radicand is greater than or equal to zero for all values of~$\theta$ in the range of interest. This implies that $c=0$. The conjugate coordinate~$\phi$ of $p_\phi$ then evolves along the deterministic drift field during the escape process, i.e., $\dot{\phi}=f_{\phi}$, and therefore~$p_{\theta}=-f_{\theta}[1\pm1]$. The $p_{\theta}=0$ solution corresponds as usual to relaxation to equilibrium, and the $p_{\theta}=-2f_{\theta}$ solution corresponds to instanton escape, antiparallel to the drift, along the $\theta$ coordinate. However, during this process the precessional dynamics of~$\phi$ remain unaltered, so escape from the stable well does not merely correspond to time-reversal of the deterministic dynamics.

Having solved for the general momenta in terms of the spherical coordinates, we can analytically derive the relation between~$\phi$ and~$\theta$ along the escape trajectories. Dividing $\dot{\phi}$ by $\dot{\theta}$ from Hamilton's equations, we find:
\begin{eqnarray}
\label{eq:phitheta}
\phi(\theta)&=&\frac{1}{\alpha}\int_0^{\theta}\mathrm{d}\tilde{\theta}\frac{\cos\tilde{\theta}}{(I+\cos\tilde{\theta})\sin\tilde{\theta}}\nonumber\\
&=&\frac{1}{\alpha(1-I^2)}\left\{\log\left[\tan(\theta/2)\right]+I\log\left[\frac{2(I+\cos\theta)}{\sin\theta}\right]\right\}\, ,\nonumber\\
\end{eqnarray}
where the integrand diverges at the $\theta=0$ pole due to degeneracy
in the~$\phi$ coordinate, and at the $\theta=\arccos(-I)$ separatrix where the escape trajectory is nondifferentiable. 

We tested these results and predictions by numerically integrating the FW~dynamics associated with with the stochastic process~(16) for several different cases. The case $D\ne 0$ is discussed in the text; here we show results for $D=0$, with both zero and small tilt.

\begin{figure}[h]
	\begin{center}
	\centerline{\includegraphics [clip,width=70mm, angle=0]{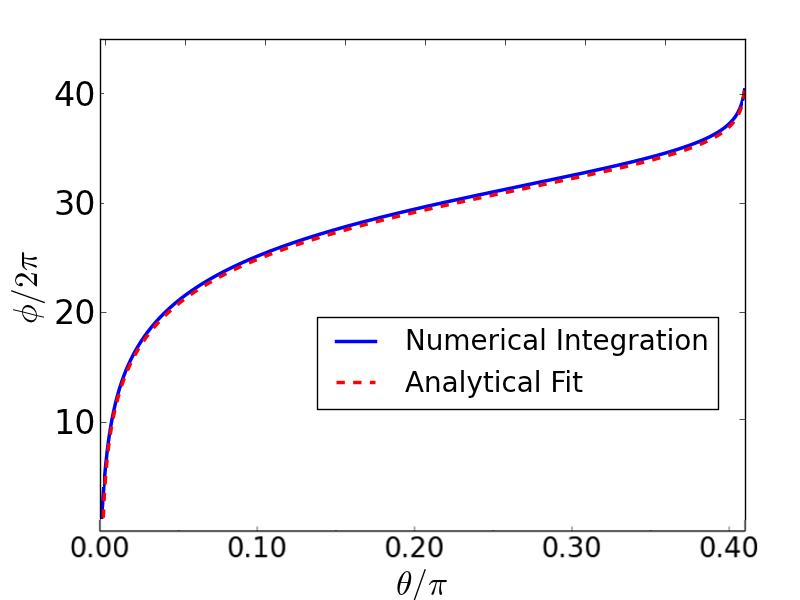}}
	\vskip -4mm
	\caption{Escape trajectory for a macrospin model with $\alpha=0.01$, $D=0$, $I=0.3\,I_C$ and $\omega=0$. Blue line shows the least-action result of numerical integration of the FW~dynamics. Dashed red line is the analytical result~(14).}
	\label{fig:uniaxial1}
	\vskip -2mm
	\end{center}
\end{figure}

Fig.~\ref{fig:uniaxial1} 
shows the optimal escape trajectory for $D=0$, $\omega=0$, and $I=0.3\,I_C$, where, as noted in the text, $I_C$ is the critical applied current above which the entire $m_z>0$ region 
becomes unstable, eliminating the bistability of the magnetic system. As shown above,, the FW~dynamics are exactly solvable in this case. 

\begin{figure}[h]
	\begin{center}
	\vskip -4mm
	\centerline{\includegraphics [clip,width=70mm, angle=0]{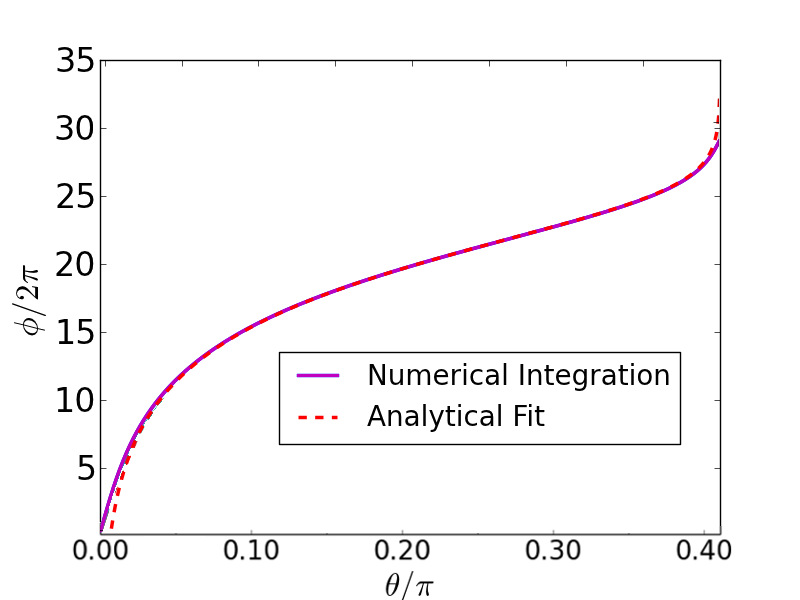}}
	\caption{Escape trajectory for a macrospin model with $\alpha=0.01$, $D=0$, $I=0.3\,I_C$ and $\omega=0.3\,\theta_C$ (defined in text). The purple line shows the least-action result of numerical integration of the FW~dynamics. The dashed red line is the analytical result~(14).}
	\label{fig:uniaxial2}
	\vskip -2mm
	\end{center}
\end{figure}

We next consider the case of small tilt. In accord with our analysis, the numerical results demonstrate that the escape dynamics are essentially unaltered~(Fig.~\ref{fig:uniaxial2}). In the limit of sufficiently small~$\alpha$, critical currents scale trivially with the tilt~($I_C^{\omega}= I_C^0/\cos\omega$)~\cite{Pinna2013}. By including this correction to the analytical expression for the $\omega=0$ escape model, we can fit the data well,  even though the system no longer obeys detailed balance. 
